\documentclass[12pt]{amsart}
\usepackage{epsfig,amssymb,amsfonts}
\usepackage[english]{babel}
\usepackage{fancyheadings} 
\newcommand{\bds}[1]{\boldsymbol{#1}}
\def\half{\mbox{\small $\frac{1}{2}$}}

\def\p{{\mathfrak p}}                            

\newcommand{\T}{\mathbb{T}}
\newcommand{\U}{\mathcal{U}}
\newcommand{\M}{\mathcal{M}}
\newcommand{\N}{\mathcal{N}}
\newcommand{\R}{\mathbb{R}}
\newcommand{\Z}{\mathbb{Z}}
\newcommand{\C}{\mathbb{C}}
\def\A{\bds A}
\def\B{\bds B}

\begin{document}
\begin{titlepage}
\begin{flushright}
     UPRF00-12\\
	July 2000 \\
\end{flushright}
\par \vskip 10mm
\begin{center}
{\large \bf Landau levels on a torus\footnote{Research supported by
 Italian MURST under contract 9702213582 and by I.N.F.N. under {\sl
 i.s. PR11}. }}
\end{center}
\par \vskip 2mm
\begin{center}
E.\ Onofri\footnote{E-mail: {\tt onofri@pr.infn.it}}\\
{\sl Dipartimento di Fisica, Universit\`a di Parma} \\
and {\sl INFN, Gruppo Collegato di Parma, 43100 Parma, Italy}

\end{center}
\par \vskip 2mm
\begin{center} { \large \bf Abstract} 
 \end{center}
\begin{quote}
{\sl Landau levels have represented a very rich field of research,
which has gained widespread attention after their application to
quantum Hall effect.  In a particular gauge, the holomorphic gauge,
they give a physical implementation of Bargmann's Hilbert space of
entire functions. They have also been recognized as a natural bridge
between Feynman's path integral and Geometric Quantization. We discuss
here some mathematical subtleties involved in the formulation of the
problem when one tries to study quantum mechanics on a finite strip of
sides $L_1, L_2$ with a uniform magnetic field and periodic boundary
conditions. There is an apparent paradox here: infinitesimal
translations should be associated to canonical operators
$[\p_x,\p_y]\propto i\hslash B$, and, at the same time, live in a
Landau level of finite dimension $B\,L_1L_2/(hc/e)$, which is impossible
from Wintner's theorem. The paper shows the way out of this conundrum.
}
\end{quote}
\end{titlepage}
\newpage

\section{Introduction}
Landau levels have been introduced in 1930 (see \cite{Landau3}).  They
found an important physical application only quite recently, after the
discovery of the Quantum Hall Effect (see \cite{Aoki,fubini,Ferrari}
and references therein).  More recently it has been recognized that
the theory of Landau levels provides a general bridge between Feynman
path integrals and ``geometric quantization'' in all cases where the
classical phase space is equipped with a complex structure which makes
it a Kaehler manifold (see \cite{KlauderEO}). From this general
viewpoint, or to get a more realistic description of conducting thin
films, it is important to understand the case of a finite region with
suitable boundary conditions. If these correspond to a compact
(smooth) manifold without boundary the quantization condition of
Kostant-Souriau (see \cite{simms} and references therein) or,
equivalently, Dirac's quantization condition for monopole charges
require that the total magnetic flux be quantized, i.e.  it must be an
integral multiple $N$ of the universal constant $hc/e$. At the same
time, the degeneracy of the ground state is finite and coincides with
$N$, except for a topological correction (half the Euler
characteristic of the manifold).  A similar, approximate, result can
be obtained by a semi-classical argument (see \cite{Landau3}).

Consider now the simple case of a rectangular area with sides $L_1,
L_2$ and periodic boundary conditions; the problem is formulated on a
toroidal surface with a transverse magnetic field, whose flux is
$BL_1L_2$. Of course this fact implies the presence of magnetic
charges, hence Dirac's quantization. The problem is: what is the
symmetry of the Hamiltonian?  We expect that the classical symmetry of
the torus ($S^1\times S^1$) be realized as a projective
representation, the two infinitesimal generators satisfying Heisenberg
algebra with a central charge $\propto \hslash B$ (at least this is
what happens in the non compact $R^2$ case). But this is clearly
incompatible (Wintner's theorem) with finite degeneracy of energy
levels! While we cannot expect a spontaneous symmetry breaking in a
system with a finite number of degrees of freedom, we know from
Geometric Quantization that {\sl not all} classical symmetries survive
at the quantum level, only those which are lifted at the pre-quantum
level and, secondly, respect the polarization (in Landau level
language, those symmetries are preserved which leave the first Landau
level invariant). The problem is: what exactly is happening on the
torus?

  To get an answer, we shall reconstruct the explicit form of the
Landau levels in terms of sections of the hermitian line bundle
associated to the principal bundle with connection given by the
magnetic potential $\A$.  The language of fibre bundles is the natural
one to describe gauge fields and it is becoming more familiar to
physicists especially after the advent of modern string theory. We
shall explicitly construct the transition functions of the line bundle
and find a natural orthonormal basis of holomorphic sections, which
turn out to be Jacobi-$\theta$-functions. By inspection it turns out that
translation invariance is broken to a discrete subgroup $Z_N\times
Z_N$, $N$ being the monopole charge. This fact has the counterpart
that the Hermitian operators which correspond to infinitesimal
translations (in the non--compact case) {\sl do not leave the Landau
levels invariant}, i.e. {\sl they do not commute with the
Hamiltonian}: while formally commuting with the Hamiltonian {as
\sl differential operators}, they fail to respect the boundary
conditions, given by the bundle transition functions.

\section{Magnetic field on the torus.}
Let $\T^2=\R^2/\Z^2$ denote the two-torus; we describe it in physical
terms by identifying an atlas of four local charts specified as
follows:
\begin{enumerate}
\item $\U_\alpha = \{0<x<L_1,\; 0<y<L_2\}$
\item $\U_\beta = \{\bar{x}<x<L_1+\bar{x},\; 0<y<L_2\}$
\item $\U_\gamma = \{0<x<L_1,\; \bar{y}<y<L_2+\bar{y}\}$
\item $\U_\delta = \{\bar{x}<x<L_1+\bar{x},\; \bar{y}<y<L_2+\bar{y}\}$
\end{enumerate}
with some choice of constants $\bar x$ and $\bar y$.  A uniform
magnetic field $B$ transverse to the surface $\T^2$ is represented by
the translation invariant two-form $\B= B\; dx \wedge dy $. The
connection form $\A$, representing the magnetic potential, is defined
in each local chart in such a way that $d\A = \B$.  It is well--known
that a global one-form on $\T^2$ satisfying this condition does {\sl
not} exist, since otherwise $\int_{\T^2} \B = \int_{\T^2} d\A = 0$, by
Stokes theorem, while it holds $\int_{\T^2} \B = BL_1L_2$. The problem
is essentially the same as the presence of a Dirac string in the case
of a three-dimensional magnetic monopole.  For the sake of simplicity,
we may define the local connection forms by the same formula $\A =
\half B (x dy - y dx)$, since the coordinates $x,y$ are indeed
differentiable within each local chart
\footnote{The correct mathematical language to describe such a setup
is that of algebraic geometry; a nice introduction for physicists can
be found for instance in \cite{alvarez85}. In this paper we try to
keep the mathematical jargon to a minimum.}. To characterize the
connection form completely we have to identify the {\sl transition
functions} which relate $\A_i$ to $\A_j$ for any pair $(i,j)$ in the
set $\{\alpha,\beta,\gamma,\delta\}$ and {\sl for each connected
component} of the overlap $\U_i\cap\U_j$; we have
\begin{eqnarray*}
\A_\beta(x,y) &=& \A_\alpha(x,y) 
\qquad\qquad\qquad\qquad\qquad (\bar x<x<L_1)\\
\A_\gamma(x,y) &=& \A_\alpha(x,y) 
\qquad\qquad\qquad\qquad\qquad (\bar y<y<L_2)\\
\A_\delta(x,y) &=& \A_\alpha(x,y) 
\qquad\qquad\qquad\qquad\qquad (\bar x<x<L_1, \bar y<y<L_2 )\\
\A_\beta(x+L_1,y) &=& 
 \A_\alpha(x,y) + d\left(\half B L_1 y + \varphi_{\alpha\beta}\right)
\qquad (0<x<\bar x)\\
\A_\gamma(x,y+L_2) &=& 
 \A_\alpha(x,y) + d\left(-\half B L_2 x + \varphi_{\alpha\gamma}\right)
\qquad (0<y<\bar y)\\
\A_\delta(x+L_1,y+L_2) &=& 
 \A_\alpha(x,y) + d\left(\half B (L_1 y - L_2 x) + \varphi_{\alpha\delta}\right) \qquad (0<x<\bar x, 0<y<\bar y)
\end{eqnarray*}
and similar transition functions for the other cases.  The constants
$\varphi_{ij}$ are arbitrary at this level; they will however play a
crucial role in the {\sl lifting} to the associated line bundle which
describes the quantum wave functions\footnote{The constants
$\varphi_{\alpha\beta}$ are connected to the fundamental cocycle
$c_{\alpha\beta\gamma}$ of Ref.\cite{simms,alvarez85}.}.
\section{The holomorphic gauge}
We now make a gauge transformation to a special gauge which is particularly convenient in the quantization process. Let us introduce complex coordinates
$z=x+iy, \bar z = x-iy$. The magnetic potential is given by
\begin{equation}\label{gauge-transf}
\A(z,\bar z) = \frac1{2i}B\, \bar z dz - \frac1{4i}B\, d |z|^2
\end{equation}
which shows that by a gauge transformation we can adopt a 
holomorphic form
\begin{equation*}
\A^h = \frac1{2i}B \,\bar z dz\;,
\end{equation*}
for which we have the transition functions
\begin{eqnarray*}
\A^h_\beta(z) &=& \A^h_\alpha(z) 
\qquad\qquad\qquad\qquad\qquad (\bar x<x<L_1)\\
\A^h_\gamma(z) &=& \A^h_\alpha(z) 
\qquad\qquad\qquad\qquad\qquad (\bar y<y<L_2)\\
\A^h_\delta(z) &=& \A^h_\alpha(z) 
\qquad\qquad\qquad\qquad\qquad (\bar x<x<L_1, \bar y<y<L_2 )\\
\A^h_\beta(z+L_1) &=& 
 \A^h_\alpha(z) + d\left(-i\half B L_1 z + \varphi_{\alpha\beta}\right)
\qquad (0<x<\bar x)\\
\A^h_\gamma(z+iL_2) &=& 
 \A^h_\alpha(z) + d\left(-\half B L_2 z + \varphi_{\alpha\gamma}\right)
\qquad\qquad (0<y<\bar y)\\
\A^h_\delta(z+L_1+iL_2) &=& 
 \A^h_\alpha(z) + d\left(-i\half B (L_1 -i L_2)z +
 \varphi_{\alpha\delta}\right) \qquad (0<x<\bar x, 0<y<\bar y)
\end{eqnarray*}
with some new choice of constants $\varphi_{ij}$.

\section{Quantization} 
The Hamiltonian for a charged particle is given by the 
minimal-coupling prescription.
The local expression as a differential operator must be complemented
by suitable boundary conditions which ensure selfadjointness.  This is
easily done in terms of a line bundle {\sl associated} to $\A$ as
defined in the previous section. The physical principle to adopt is
the gauge principle, according to which 
\begin{equation*}
\left(i\hslash\partial_\mu -\frac{e}{c} A_\mu\right)\,\psi
\end{equation*}
is covariant under gauge transformations, in particular under the
transition from one chart to another (here $A_\mu$ are the components
of the gauge potential one-form $\A=\sum A_\mu dx^\mu$).  This can be
done directly in the holomorphic gauge, which is our choice for the
sequel. As usual, the complex line bundle has transition functions
obtained by exponentiating those which characterize $\A^h$:
\begin{eqnarray*}
\psi_\beta(z) &=& \psi_\alpha(z) \qquad\qquad (\bar x<x<L_1)\\
\psi_\gamma(z) &=& \psi_\alpha(z) \qquad\qquad (\bar y<y<L_2)\\
\psi_\delta(z) &=& \psi_\alpha(z) \qquad\qquad (\bar x<x<L_1, \bar
y<y<L_2 )\\ 
\psi_\beta(z+L_1) &=& \psi_\alpha(z)\;
\exp\left\{\frac{eBL_1}{2\hslash c} z + \phi_{\alpha\beta}\right\}
\qquad\qquad\quad (0<x<\bar x)\\ 
\psi_\gamma(z+iL_2) &=&
\psi_\alpha(z)\;\exp\left\{ -\frac{ieBL_2}{2\hslash c} z +
\phi_{\alpha\gamma}\right\} \qquad\qquad (0<y<\bar y)\\
\psi_\delta(z+L_1+iL_2) &=& \psi_\alpha(z)\;\exp\left\{\frac{eB(L_1
-i L_2)}{2\hslash c} z + \phi_{\alpha\delta}\right\} \qquad (0<x<\bar
x, 0<y<\bar y)
\end{eqnarray*}
where we have redefined the constants $\varphi\to\phi$ to absorb a
common factor $ieB/\hslash c$. It is clear that both
$\overline\partial\psi$ and $(\partial - z)\psi$ transform in the same
way as $\psi$.  We have to stress here that while $\phi_{ij}$ are
totally arbitrary, they {\bf must} be chosen once for all to define
the Hamiltonian; as we shall show, different choices correspond in
general to unitarily equivalent, yet distinct, operators. The
situation is rather different from the well-known Aharonov-Bohm case,
where the various admissible boundary conditions yield inequivalent
Hamiltonians.
 
The local expression of the Hamiltonian in terms of complex
coordinates is easily found to be
\begin{equation}\label{eq:2}
H^h = -4\frac{\hslash^2}{2m}\left(\partial - \frac{eB}{2mc\hslash}\bar z\right) \bar\partial
\end{equation}
($\partial\equiv\partial/\partial z,\,\bar\partial\equiv\partial/\partial \bar z$)
where we dropped a zero point energy term $\hslash\omega$.

We must now introduce the Hermitian structure which allows to define
the quantum inner product between wave functions. It is readily seen
(e.g. starting from the Euclidean inner product in the real gauge and
performing the gauge transformation to the holomorphic case) that the
Hermitian structure is given by
\begin{equation*}
h(\psi_1,\psi_2) =
\exp\{-\frac{eB}{2\hslash c}|z|^2\}\; \overline{\psi_1}\,\psi_2\;.
\end{equation*}
in terms of which we can define the quantum inner product
\begin{equation*}
\langle\psi_1|\psi_2\rangle = \int_{\T^2} h(\psi_1,\psi_2)\;
[dz]
\end{equation*}
where $[dz]\equiv\frac1{2i}\overline {dz} \wedge dz $.  It is easy to
check that there is a smooth match $h(\psi_i,\psi_i) =
h(\psi_j,\psi_j)$ on each $\U_i\cap\U_j$ provided that
$\Re(\phi_{ij})$ be suitably chosen.  To simplify the notation, let us
introduce natural units adapted to the problem: let us use
$(\hslash/m\omega)^{\half}$ as length unit, where $\omega=eB/2mc$ is
Larmor's frequency. Then we get the new transition functions which
make $h(\psi,\phi)$ smooth. At this point we can drop the chart index
from the wave-function: from our convention, there is an open set
common to all local chart where the wave function is the same in all
local charts and the transition functions merely represent the
boundary conditions to be imposed on $\psi$.

\begin{equation}\label{eq:1}
\begin{split}
\psi(z+L_1) &= \psi(z)\; \exp\left\{L_1 z + \half L_1^2 +
i\delta_1\right\}\\ 
\psi(z+iL_2) &= \psi(z)\;\exp\left\{ -i L_2 z + \half L_2^2 + i
\delta_2 \right\} 
\end{split}
\end{equation}

It is also easily checked that these b.c. make the Hamiltonian
hermitian. ({\small Hint: make use of the complex integration by parts in the form 
$
\int_{\T^2} \overline{dz}\wedge dz \;\overline{\phi(z)}\,
\overline\partial \psi(z) = \int_{\T^2} \overline{\phi(z)}\, d
\left(\psi dz\right) =\\ 
\oint \overline{\phi(z)}\,\psi(z)\,dz -
\int_{\T^2} \overline{dz}\wedge dz \;\overline{\partial\phi(z)}\, \psi$
}).

However, there is a consistency condition to be satisfied,
which stems from a general theorem about hermitian line bundles due to
A.Weil (see \cite{weil}; a simple proof taken from \cite{simms} is
reproduced in Appendix \ref{sec:diracs-quantization}). In our case it
can be found as follows: by successively applying the previous
relations we get

\begin{equation}\label{quantization}
\begin{split}
\psi(z+L_1+iL_2) &= \psi(z+L_1)\;\exp\{-iL_2(z+L_1) +\half L_2^2 + \delta_2\}\\
	&= \psi(z)\;\exp\{(L_1-iL_2) z + \half|L_1+iL_2|^2 + i\delta_1 + i\delta_2 -iL_1L_2\}\\
	&= \psi(z+iL_2)\;\exp\{L_1(z+iL_2) +\half L_1^2 + \delta_1\}\\
	&= \psi(z)\;\exp\{(L_1-iL_2) z + \half|L_1+iL_2|^2 + i\delta_1 + i\delta_2 +iL_1L_2\}
\end{split}
\end{equation}
hence
\begin{equation*}
2L_1L_2 = 2N\pi\;,
\end{equation*}
which is Dirac-Weil-Kostant-Souriau quantization condition.
Let us conclude this section by giving the explicit expression for $\langle\psi|H|\psi\rangle$, which 
exhibits $H$ as a positive operator:
\begin{equation*}
\langle\psi|H|\psi\rangle = \int_{\T^2} [dz] e^{-|z|^2}\,|\bar\partial\psi|^2 \,,
\end{equation*}
a general result for quantum mechanics on Kaehler manifolds
\cite{KlauderEO}, from which we get the general result that the ground
state coincides with the subspace of holomorphic sections ($\bar\partial
\psi=0$).

\section{Finite--dimensional Landau levels}

We can now compute the solutions of Schroedinger equation belonging to
the ground state. These are given by holomorphic functions satisfying
the boundary conditions (\ref{eq:1}). Let us choose $\delta_1=
\delta_2=0$. Setting $\psi(z)=\exp\{\half z^2\}\,\theta(z)$, we find
that $\theta$ must be periodic with real period $L_1$ hence it can be
expanded in a Fourier series $\theta(z)=\sum_{n=-\infty}^{\infty}
c_n\,e^{2\pi i n z/L_1}$). It follows
\begin{equation*}
\begin{split}
\psi(z+iL_2) = e^{\half(z+iL_2)^2} \sum_{n=-\infty}^{\infty}
c_n\,e^{2\pi i n z/L_1}e^{-2 \pi n L_2/L_1}\\ = e^{\half
z^2}\sum_{n=-\infty}^{\infty} c_n\,e^{2\pi i n z/L_1} e^{-iL_2z +\half
L_2^2}\;.
\end{split}
\end{equation*}
which gives
\begin{equation*}
e^{2iL_2z-L_2^2} \sum_{n=-\infty}^{\infty} c_n\,e^{2\pi i n z/L_1} 
e^{-2n\pi L_2/L_1}= \sum_{n=-\infty}^{\infty} c_n\,e^{2\pi i n z/L_1} \,.
\end{equation*}
Making use of Dirac's quantization ($L_2=N\pi/L_1$) we get the
condition
\begin{equation*}
\sum_{n=-\infty}^{\infty} c_n\,e^{2\pi i (n+N)z/L_1}e^{-2n\pi L_2/L_1-L_2^2}=
\sum_{n=-\infty}^{\infty} c_n\,e^{2\pi i n z/L_1}
\end{equation*}
which is readily transformed into the recurrence relation
\begin{equation*}
c_n = c_{n-N} \,e^{-2n\pi L_2/L_1+2N\pi L_2/L_1-L_2^2}
\end{equation*}
whose solution is
\begin{equation*}
c_n = e^{-\pi n^2L_2/(L_1 N)}\; b_n
\end{equation*}
where $b_n$ is such that $b_n=b_{n+N}$. Hence {\sl there are $N$
orthogonal solutions} given by
\begin{equation*}
\left\{ \psi_\nu(z) = \N_\nu\;e^{\half z^2}\,
\sum_{n\equiv\nu\mod(N)}\exp\{-\frac{\pi n^2 L_2}{NL_1} + 2n\pi i z/L_1\}\,
\bigg|\quad \nu=0,1,\ldots,N-1\right\}\;.
\end{equation*}
We can obtain a new representation in terms Gaussian functions, very
convenient for a practical calculation of $\psi$, by applying
Poisson' summation formula (see for e.g. \cite{Lighthill}). We find
\begin{equation*}
\psi_\nu(z) = \N_\nu e^{\half z^2 + 2\pi i\nu z/L_1}
\sum_{n=-\infty}^{\infty} \exp\{-(z+n L_1/N + i\nu L_2/N)^2\}\;.
\end{equation*}
Higher levels can be simply obtained by applying the covariant
creation operator $\partial-\bar z$ to each $\psi_\nu$. 

\section{Translation symmetry breaking}
\begin{figure}[ht]
\begin{center}
\mbox{\epsfig{figure=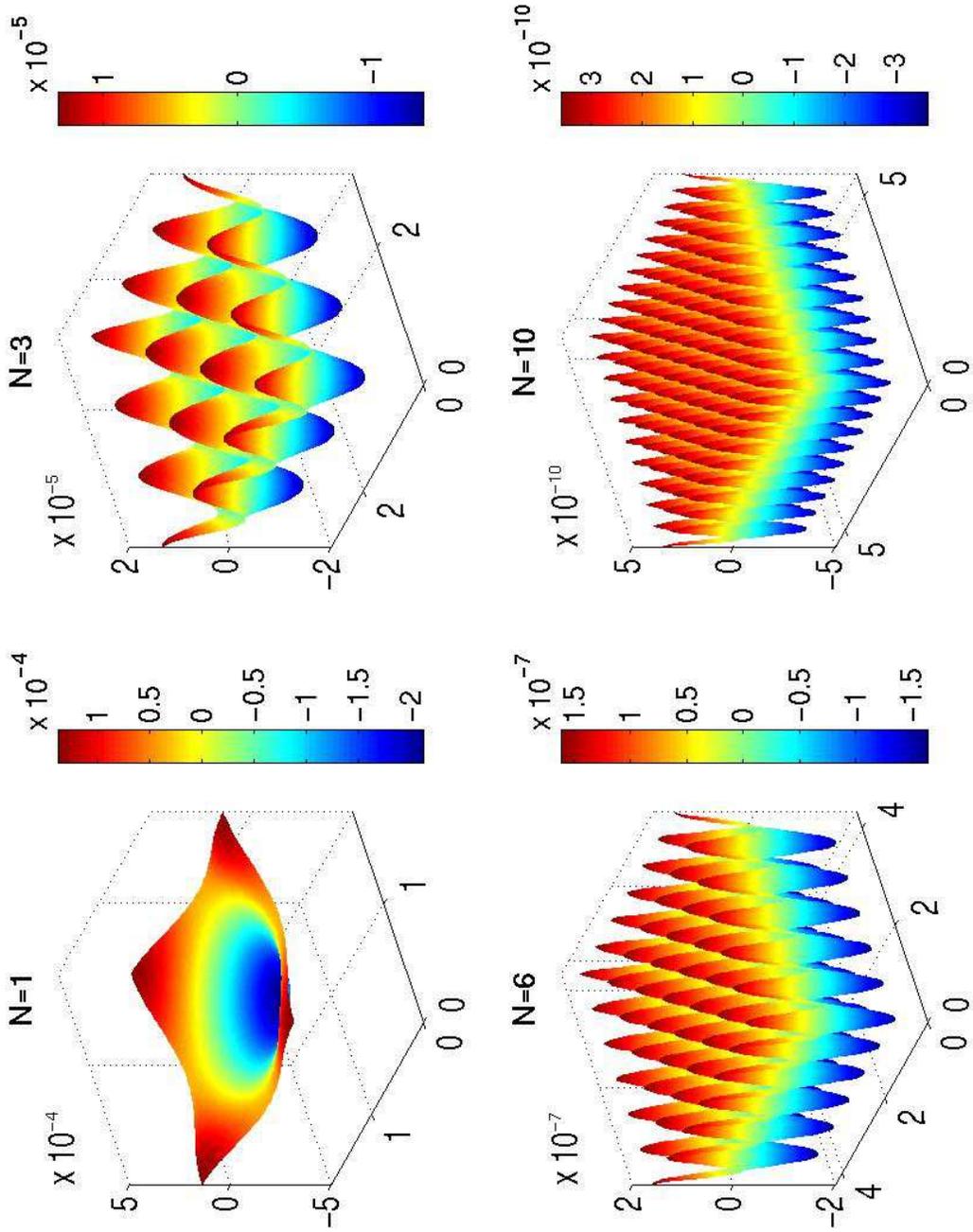,width=15.cm}}
\caption{\em Deviation from uniformity of $\rho=\Sigma_{\nu=0}^{N-1}
|\psi_\nu(z)|^2$, $N=1, 3, 6, 10$.}\label{fig:1}
\end{center}
\end{figure}

\begin{figure}[ht]
\begin{center}
\mbox{\epsfig{figure=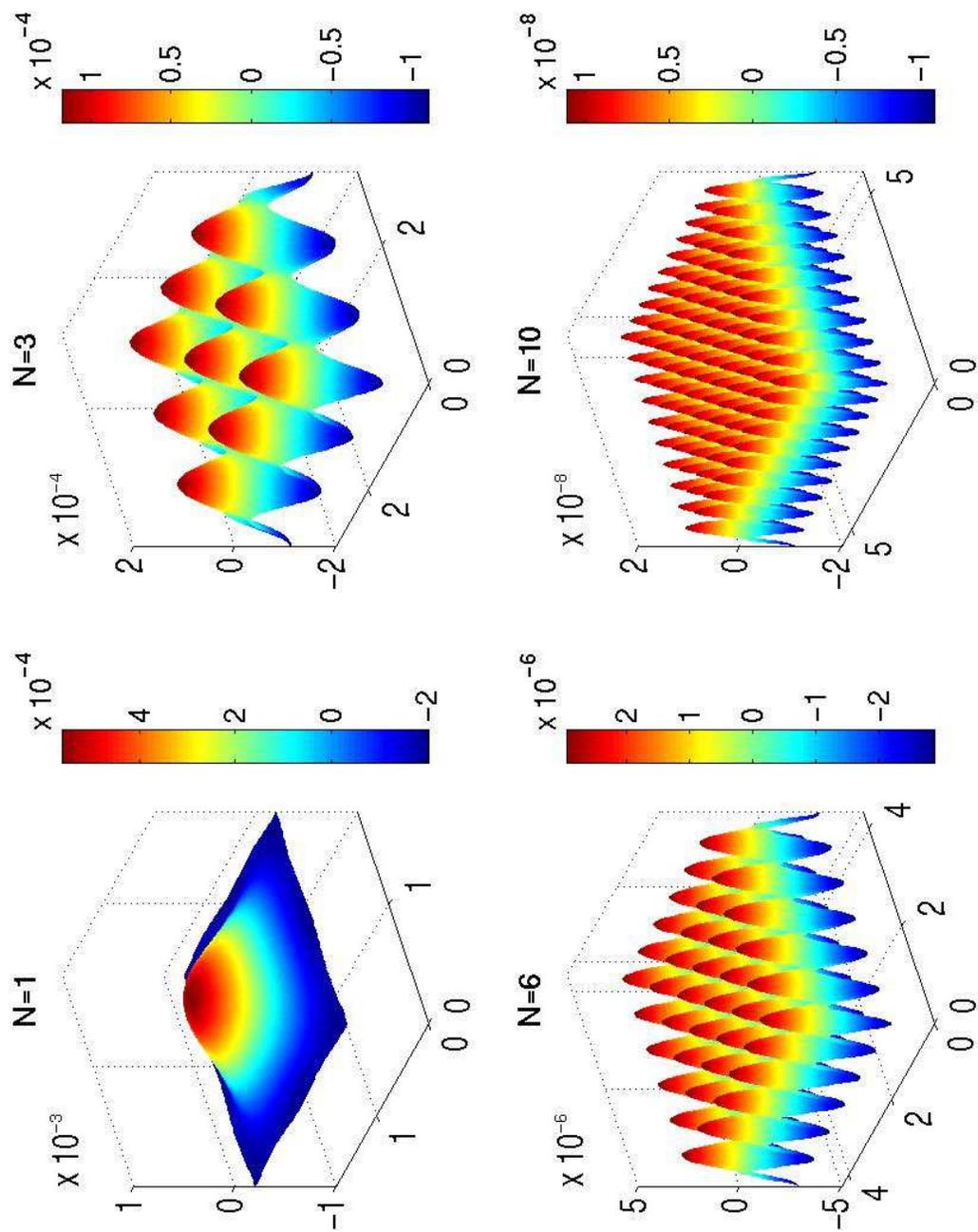,width=15.cm}}
\caption{\em Deviation from uniformity of $\rho$ in the second
Landau level, $N=1, 3, 6, 10$.}\label{fig:2}
\end{center}
\end{figure}

The main question which started this investigation was the following:
what happens to the translation symmetry of the torus? The question is
motivated from the fact that unitary translations are realized as
projective representations with a ``central charge'' given by the
magnetic field strength. Hence they cannot live in a finite
dimensional space (see Appendix \ref{sec:group-theor-wintn}).  Before
going to analyze the problem in great detail, just observe that under
the assumption that such a translation symmetry would nevertheless
survive in some way, we should see it as a property of the ground
state, i.e.  there must exist a finite unitary matrix $t_{\mu\nu}$
such that $(T_a\,\psi_\nu)(z) = \sum_{\mu}\,t_{\nu\mu}(a)\psi_\mu(z)
$. It would follow that the density matrix
$\rho_N(z)=\sum_\nu|\psi_\nu(z)|^2$ should then be translation
invariant, i.e. constant on the torus.  If we calculate $\rho_N$ for
the first few values of $N$ we immediately find that this is not
so. The density $\rho$ exhibits a series of regularly spaced bumps,
precisely at the location $(n_1 L_1 + n_2L_2)/N$ (see Fig.s
\ref{fig:1}-\ref{fig:2}, where the deviation from uniformity is
plotted for the first two Landau levels at various values of the
magnetic charge).

As is clear from the pictures, translation symmetry is broken, presumably to
$Z_N\times Z_N$, but the breaking tends to be weaker at high $N$
(a variation of $O(10^{-N})$). Is there a simple explanation of this
symmetry breaking? 
The point is that we can easily implement compact translations in the same way as we can do in the non-compact case. The unitary operators are given by
\begin{equation*}
(T_a\,\psi)(z) = e^{\bar a z-\half |a|^2}\,\psi(z-a)
\end{equation*}
where the value of $\psi$ should be found through the twisted
periodicity conditions given in Eq.~(\ref{eq:1}). It is readily checked that
\begin{enumerate}
\item $T_a$ {\sl formally} commute with the Hamiltonian, i.e. with the
differential operator of Eq.~(\ref{eq:2});
\item $T_aT_bT_{-a}T_{-b}=\exp\{\bar a b - a \bar b\}$;
\item $T_a$ does {\sl not} in general leave the ground state invariant,
i.e. invariance is maintained only if  $Na$ is trivial, that is
$a=(n_1L_1+in_2L_2)/N$;
\item the formal infinitesimal generators of $T_a$, namely
$\p_1=iz-i(\partial+\overline\partial)$ and
$\p_2=-iz-i(\partial-\overline\partial)$ {\bf do not} leave the space
of sections (Eq.~\ref{eq:1}) invariant.
\end{enumerate}
To begin with the last statement, it is clear that we may consider the
linear combinations $\partial$ and $z-\overline\partial$, neither of
which is such as to transform sections into sections.  From the group
point of view, let $\ell$ be a translation in $\Z_2$, i.e.
$\ell=k_1L_1+ ik_2L_2, k\in\Z$. Let us consider $T_a\psi$. We find
\begin{equation*}
\begin{split}
(T_a\psi)(z+\ell) &= \exp\{\bar a(z+\ell)-\half|a|^2\}\,
\exp\{\bar\ell (z-a) +\half|\ell|^2\}
\psi(z-a)\\
&=(T_a\psi)(z)\,\exp\{\bar\ell z +\half|\ell|^2 + \bar a \ell -a\bar\ell\}\,.
\end{split}
\end{equation*}
We conclude that a translated section satisfies boundary conditions with a different choice of the constants $\delta_1, \delta_2$, hence the bundle structure is not invariant under translation, except for
\begin{equation*}
\bar a \ell - a \bar \ell = 2i\Im\{\bar a \ell\} \in 2\pi i \Z
\end{equation*}
which occurs precisely when $a=(n_1L_1+in_2L_2)/N$ ($\Im\{\bar a \ell\}
= (n_1k_2-n_2k_1)L_1L_2/N = (n_1k_2-n_2k_1)\pi$, by Dirac's
quantization).

\section{Conclusions}
The problem of a constant magnetic field transversal to a torus raises
the problem of translational symmetry. By quantizing the system
according to the standard mathematical formulation of gauge theory we
have shown that the symmetry is broken to $Z_N \times Z_N$. The
conclusion to which one is led by this result is that the ambiguity in
quantization, namely the two arbitrary phases $\delta_1, \delta_2$,
entering in the definition of the domain of the Hamiltonian operator,
represent some physical degree of freedom of the magnetic charge
distribution generating the uniform field on the torus: monopole
charges have, so to speak, horns. The effect is purely quantum
mechanical and we empirically established that it vanishes
approximately as $ \exp\{-O(B/\hslash)\} $. The mathematical roots of
the result are the classic theorems of Weil (see \cite{weil}, Ch.VI,
n.3, Prop.3); a thorough study of $\theta$-functions can be found in
\cite{dubrovin}.
\newpage
\begin{center}
{\sc Acknowledgments}
\end{center}
I thank warmly P. Maraner and C. Destri for interesting discussions
and for driving my attention to refs.\cite{fubini, dubrovin}. 

\appendix
\section{A group-theoretical Wintner's theorem}\label{sec:group-theor-wintn}
Wintner's theorem (see \cite{Putnam}) states that the identity
operator in a Hilbert space cannot be the commutator of two bounded
operators. There is a poor's man version of the theorem. Let
$U(a)$ and $V(b)$ be unitary operators satisfying the canonical commutation relations (at the group level) 
\begin{equation*}
U(a)V(b)U(-a)V(-b)=e^{\bar a b - a\bar b}\,,\qquad (a,b\in\C)\,.
\end{equation*}
Then $U$ and $V$ cannot be finite dimensional matrices. 

\noindent
{\sf Proof}: just evaluate the determinant of both sides to get 
\begin{equation*}
 1 = \exp\{2iN \Im(\bar a b)\} \quad\text{with $N=\dim(U)$\,}.
\end{equation*}
 This is a contradiction, since the r.h.s. can assume any value on the
unit circle.  This last equation shows that we may take $a$ and $b$ in
a finite subgroup and preserve the commutation relation: let
$Z_N=\{(n_1L_1+in_2L_2)/N|n_i\in\Z\}$; then the condition is satisfied
precisely if $L_1L_2=N\pi$.

\section{Dirac-Weil-Kostant-Souriau quantization condition}
\label{sec:diracs-quantization}

A general theorem (\cite{hirzebruch78}, Th.21.1) relates the dimension
of spaces of closed holomorphic forms on complex vector bundles to
geometrical objects, namely Chern and Todd classes of the base space
and of the bundle. In the simple case of a line bundle (fibre equal to
$\C$) over a complex two dimensional Riemann surface the theorem
reduces to a simple result which has a very intuitive flavour from the
point of view of Geometric Quantisation: the dimension of the physical
Hilbert space coincides with the volume of phase space in units
$\hslash$ plus a constant given by half the Euler characteristic of
the surface.  This in turn implies that the volume of phase space must
be an integer.  We report here what appears to be the simplest proof,
covering Dirac's quantization condition, combining ideas from
\cite{alvarez85} and \cite{simms}. Let us build a triangulation of the
surface $\M$ with vertices $\alpha, \beta, \gamma, ...$. Let
$\U_\alpha$ denote the union of all triangles having $\alpha$ as
vertex. By taking a sufficiently fine mesh, non empty intersections
$\U_\alpha\cap\U_\beta$ consist of the union of two triangles which
share the side $\alpha-\beta$. A gauge field on $\M$ is given by a
closed two-form $\bds B$; in each local chart $\U_\alpha$ we define a
potential $\A_\alpha$ such that $\bds B = d\A_\alpha$ in $U_\alpha$.
According to Poincar\'e Lemma, for neighboring local charts we have
\begin{equation*}
\A_\alpha - \A_\beta = d\chi_{\alpha\beta}\,,
\end{equation*}
with differentiable {\sl transition functions} $\chi_{\alpha\beta}$
which are antisymmetric in their indices. On triple intersections
(any triangle $\U_\alpha\cap\U_\beta\cap\U_\gamma$) we have
\begin{equation*}
\A_\alpha - \A_\beta = d\chi_{\alpha\beta}\,,\quad
\A_\beta - \A_\gamma = d\chi_{\beta\gamma}\,,\quad
\A_\gamma - \A_\alpha  = d\chi_{\gamma\alpha}
\end{equation*}
It follows that
$c_{\alpha\beta\gamma}\equiv\chi_{\alpha\beta}+\chi_{\beta\gamma}+\chi_{\gamma\alpha}
$ is constant on each triangle\footnote{It is useful to regard the
relation between $\A$, $\chi$ and $c$ in terms of the {\sl coboundary}
operator: $(\delta\A)_{\alpha\beta}=d\chi_{\alpha\beta}$,
$c_{\alpha\beta\gamma}= (\delta\chi)_{\alpha\beta\gamma}$. See
\cite{alvarez85}.}.  Let us introduce a line bundle associated to
$\B$: it is given locally by a direct product $\U_\alpha\times \C$ in
such a way that in any overlap the complex fibres are connected
by\footnote{In the physical application the phase is $e\chi/\hslash
c$.}
\begin{equation*}
\zeta_\alpha = \zeta_\beta\,\exp\{i\chi_{\alpha\beta}\}
\end{equation*}
For consistency, on any triple overlap it must hold
\begin{equation*}
\exp\{i\chi_{\alpha\beta}+i\chi_{\beta\gamma}+i\chi_{\gamma\alpha}\}=1
\end{equation*}
which implies thar $c_{\alpha\beta\gamma}$ must be an integer multiple
of $2\pi$.
This is usually referred as Weil theorem on holomorphic vector bundles 
(\cite{weil}, Prop.1, Ch.V, N.4). 

The key result for our purposes is the following: 

\noindent
{\sf Theorem}: the integral $\int_\M\B$ coincides with the discrete sum
$\sum_{\Delta} c_\Delta$ where $\Delta$ runs over all triangles of the mesh. 

\noindent
{\sf Proof}: 
The following purely algebraic identity holds (\cite{simms}, p.131):
\begin{equation*}
\begin{split}
\int_{\Delta_{\alpha\beta\gamma}} \B = \hfill
\frac13\oint_{\partial\Delta_{\alpha\beta\gamma}} \left(
\A_\alpha+\A_\beta+\A_\gamma\right) \hfill\\
=\frac13\left[(\chi_{\alpha\beta}+\chi_{\beta\gamma}+\chi_{\gamma\alpha})(\alpha)+
(\chi_{\alpha\beta}+\chi_{\beta\gamma}+\chi_{\gamma\alpha})(\beta) +
(\chi_{\alpha\beta}+\chi_{\beta\gamma}+\chi_{\gamma\alpha})(\gamma)\right]\\
-\frac12\left\{(\chi_{\alpha\beta}(\alpha)+\chi_{\alpha\beta}(\beta))+
(\chi_{\beta\gamma}(\beta)+\chi_{\beta\gamma}(\gamma))+
(\chi_{\gamma\alpha}(\gamma)+\chi_{\gamma\alpha}(\alpha))\right\}\\ +
\frac12\left\{ \int_{\alpha\beta}(\A_\alpha+\A_\beta)+
\int_{\beta\gamma}(\A_\beta+\A_\gamma)+
\int_{\gamma\alpha}(\A_\gamma+\A_\alpha)\right\}
\end{split}
\end{equation*}
Notice that the terms in curly brackets average to zero when we sum
over the whole triangulation, while the terms in square brackets are
precisely the cocycle $c_\Delta$, whose value is constant on the
triangle.  Hence we get
\begin{equation*}
\int_\M\B = \sum_{\Delta}c_{\Delta}
\end{equation*}
and as a result the flux of $\B$ is quantized.

\bibliographystyle{amsalpha}

\bibliography{mybiblio} 


\end{document}